\documentclass[preprint,12pt,3p]{elsarticle}

\usepackage{graphicx}

\usepackage{amssymb}

\usepackage{lineno}

\usepackage{soul,color}

\usepackage[hidelinks]{hyperref}

\usepackage{subcaption}

\journal{Nuclear Instruments and Methods in Physics Research}

\begin{document}

\begin{frontmatter}

\title{On Determining Dead Layer and Detector Thicknesses for a Position-Sensitive Silicon Detector}

\author[label1]{J. Manfredi\corref{cor1}}
\ead{manfredi@nscl.msu.edu}
\cortext[cor1]{Corresponding author}
\author[label2]{Jenny Lee}
\author[label1]{W.G. Lynch}
\author[label1,label3]{C.Y. Niu}
\author[label1]{M.B. Tsang}
\ead{tsang@nscl.msu.edu}
\author[label1]{C. Anderson}
\author[label1]{J. Barney}
\author[label1]{K.W. Brown}
\author[label4]{Z. Chajecki}
\author[label1]{K.P. Chan}
\author[label4]{G. Chen}
\author[label1]{J. Estee}
\author[label3]{Z. Li}
\author[label5]{C. Pruitt}
\author[label6]{A.M. Rogers}
\author[label1]{A. Sanetullaev}
\author[label1]{H. Setiawan}
\author[label1]{R. Showalter}
\author[label1]{C.Y. Tsang}
\author[label1]{J.R. Winkelbauer}
\author[label7]{Z. Xiao}
\author[label2]{Z. Xu}

\address[label1]{NSCL and Department of Physics and Astronomy, Michigan State University, East Lansing, MI 48824, USA}
\address[label2]{Department of Physics, The University of Hong Kong, Pokfulam Road, Hong Kong, China}
\address[label3]{School of Physics and State Key Laboratory of Nuclear Physics and Technology, Peking University, Beijing 100871, China}
\address[label4]{Department of Physics, Western Michigan University, Kalamazoo, MI 49008, USA}
\address[label5]{Department of Chemistry, Washington University, St. Louis, MO 63130, USA}
\address[label6]{Department of Physics, University of Massachusetts Lowell, Lowell, MA 01854, USA}
\address[label7]{Department of Physics, Tsinghua University, Beijing 100084, China}

\begin{abstract}
In this work, two particular properties of the position-sensitive, thick silicon detectors (known as the ``E'' detectors) in the High Resolution Array (HiRA) are investigated: the thickness of the dead layer on the front of the detector, and the overall thickness of the detector itself. The dead layer thickness for each E detector in HiRA is extracted using a measurement of alpha particles emitted from a $^{212}$Pb pin source placed close to the detector surface. This procedure also allows for energy calibrations of the E detectors, which are otherwise inaccessible for alpha source calibration as each one is sandwiched between two other detectors. The E detector thickness is obtained from a combination of elastically scattered protons and an energy-loss calculation method. Results from these analyses agree with values provided by the manufacturer. 
\end{abstract}


\end{frontmatter}


\section{Introduction}
\label{intro}

Since the 1950s, semiconductor materials have been commonly used in radiation detection devices \cite{Mckay51,McKenzie59,Brom60,Miller60}. For charged-particle detection, silicon is the most popular choice of semiconductor. During fabrication of a silicon detector, a “dead layer” typically forms on the surface to protect the active silicon wafer of the detector \cite{Knoll10}. Energy deposited by charged particles in this dead layer is not detected. Therefore the dead layer affects the accuracy of charged-particle energy measurements. Silicon dead layers have previously been measured using low energy electrons \cite{Wall13} and proton bremsstrahlung \cite{Cohen08}. Another important characteristic of a silicon detector is its overall thickness, which determines the ``punch-through energy'' required for a particle to pass completely through the detector (often to additional detectors located behind). Understanding both these characteristics is necessary to properly use the detectors for accurate and precise energy measurements of charged particles.

\begin{figure}
\centering
\includegraphics[width=0.7\textwidth]{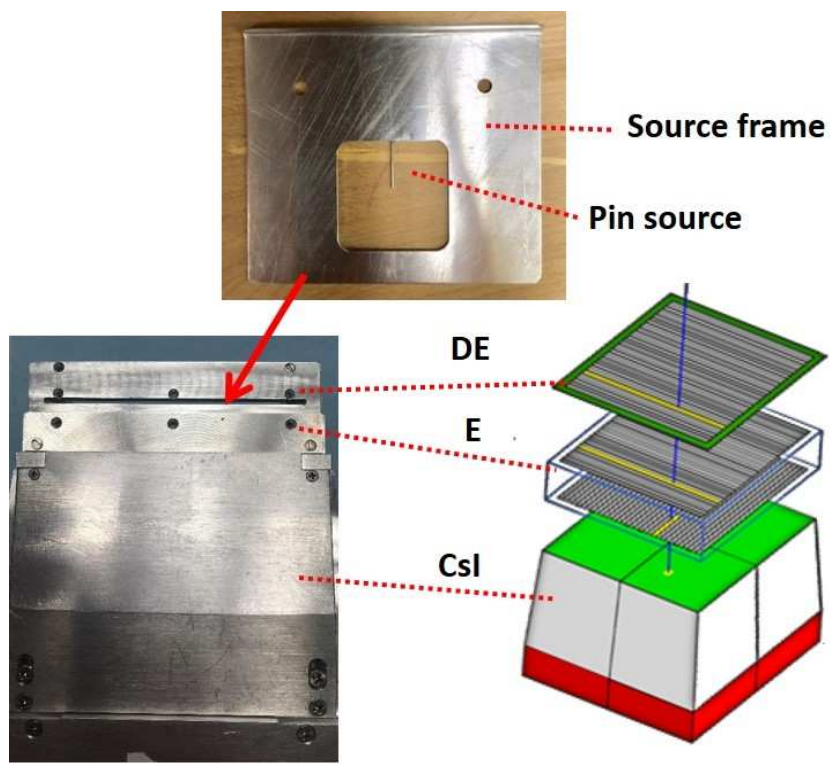}
\caption{HiRA telescope assembly. On the bottom left is a side-view photograph of a HiRA telescope. On the bottom right is a cartoon of the detectors contained within the telescope. Dotted red lines connect the detectors in the cartoon to their approximate positions in the telescope. On the top of the figure is the pin source frame with the pin source installed. The pin source slot is located between the DE and the E detector as indicated by the thick red arrow.}
\label{fig:hiralayout}
\end{figure}

Silicon detectors are integral parts of many actively used arrays in nuclear physics, including LASSA \cite{Davin01}, MUST2 \cite{Poll05}, ANASEN \cite{Kosh17}, and HELIOS \cite{Wuos07}. The High Resolution Array (HiRA) \cite{Wallace07} is another such example. Each HiRA ``telescope'' consists of a 65-$\mu$m single-sided silicon strip detector (known as the ``DE''), a 1500-$\mu$m double-sided silicon strip detector (``E''), and an array of 4 CsI scintillator crystals read-out by silicon photodiodes. Both the E and DE silicon detectors are manufactured by Micron Semiconductor \cite{MicronAddress}. 20 HiRA telescopes have been built allowing the array to be configured differently to optimize the study of different physics objectives in each experiment. These components are enclosed in a metal can, shown in the left of Figure \ref{fig:hiralayout}. A slot in the side of the metal can between the DE and E accommodates a calibration source (known as the pin source) that will be discussed in detail below. At the front of each telescope is a thin Mylar foil that protects the sensitive silicon detectors and also completes a Faraday cage around the detectors to minimize noise. The strips on opposite sides of the E detector are perpendicular to each other, so detecting a particle in both a front strip and a back strip results in a single 1.95 mm by 1.95 mm ``pixel''. The small size of each pixel yields excellent position resolution. Furthermore, by correlating the energies deposited in different detectors within a telescope, HiRA provides particle identification information for a wide range of isotopes and energies. Because of these two features, HiRA is a powerful tool for studying a wide variety of topics in nuclear physics, including direct reactions for nuclear structure, nuclear astrophysics, exotic two-proton decay, and the nuclear symmetry energy \cite{Wuosmaa17,Rogers11,Brown14,Henzl12}.

Although each HiRA telescope contains two different silicon detectors, the characterizations presented here pertain only to the E detectors. The organization of this work is as follows: first, the experimental features and analysis for determination of the dead layer thickness on the front of each E detector in HiRA will be presented. Next follows the experimental features and analysis performed to extract the overall detector thickness of each E detector. Finally, the results will be summarized.

\section{Dead Layer Thickness Determination}
\subsection{Experimental Details}
\begin{figure}
\centering
\includegraphics[width=0.7\textwidth]{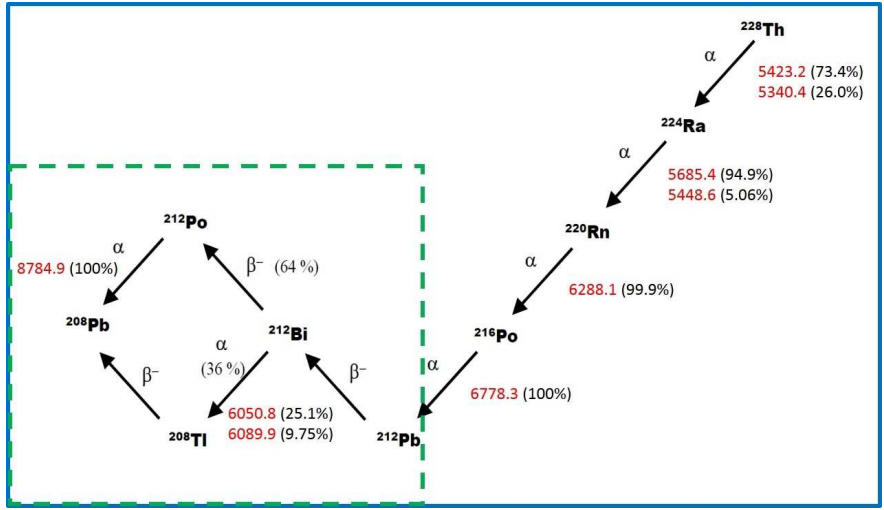}
\caption{Decay scheme for $^{228}$Th. The green dashed box includes all decay radiation from $^{212}$Pb, the isotope that is primarily deposited onto each pin source. Energies for all alpha decays with branching greater than 1\% are shown in red, along with the corresponding branching rates \cite{ENSDF}.}
\label{fig:decays}
\end{figure}

\begin{figure}
\centering
\includegraphics[width=0.7\textwidth]{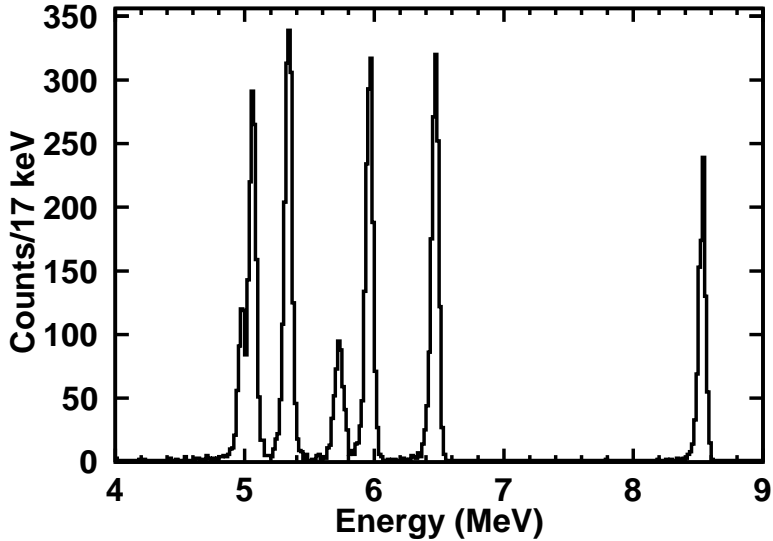}
\caption{$^{228}$Th alpha spectrum for an example HiRA E detector. Energy losses in the gold window, the Mylar foil in the front of the HiRA telescope, and the E dead layer are taken into account. Typically, a HiRA silicon calibration will use the 5 most prominent peaks seen in the spectrum. The resolution in this detector is 65 keV (FWHM).}
\label{fig:exampleTh}
\end{figure}

A $^{228}$Th alpha source is commonly used to calibrate silicon detectors and is the calibration source of choice for the HiRA E and DE detectors. Using a $^{228}$Th source has three key advantages. First, there are six clearly separated peaks with energies from about 5 MeV to about 9 MeV. Second, $^{228}$Th sources of various strengths are commercially available. Lastly, $^{228}$Th has a relatively long half-life (1.9 years). The source used in the current work was electroplated onto a platinum surface, and then fixed in an aluminum holder (12.7 millimeters in diameter and 6.35 millimeters tall) with a 100 $\mu$g/cm$^2$ gold window. Figure \ref{fig:decays} shows the decay radiation of $^{228}$Th and its daughters \cite{ENSDF}, and Figure \ref{fig:exampleTh} shows an example energy spectrum for an E detector in which the peaks from these decays can be clearly seen. In HiRA calibrations, typically the five largest peaks are used.

Because the DE detector blocks alpha particles from passing through to the E, calibrating the E with a $^{228}$Th source requires removing the DEs from all telescopes. This can only be done by disassembling the entire array, removing the DE detectors, and then reassembling the array. Since energy calibrations can be sensitive to minor changes in electronics and cable configurations, it is necessary to confirm that the performance of the E detector after reassembling the array is consistent with the performance of the E prior to disassembly. Ideally, if an alpha source can be inserted between the DE and E detectors, the E detector can be calibrated without disassembling the HiRA telescope. 

To resolve this issue, each HiRA telescope can features a slot between the DE and the E detectors for insertion of a pin source, which is made from a small metal pin. One end of the pin is covered with isotopes from the decay chain of $^{228}$Th (primarily $^{212}$Pb). To make these pin sources, a negative bias is applied to a bundle of pins placed in front of a 13 $\mu$Ci $^{228}$Th powder source. When $^{228}$Th decays to $^{220}$Rn, which has a relatively long half-life (56 seconds), some of the gaseuos $^{220}$Rn ions drift to the negatively-charged pin electrodes. The $^{220}$Rn then decays into the short-lived $^{216}$Po and the much longer-lived $^{212}$Pb (0.1 seconds and 10.6 hours, respectively), which are metals and generally stick to the head of the pin. Typically 20 pins are irradiated at a time, and after about 24 hours of irradiation, the pins are each mounted on a frame that is then inserted into the aligned slot between the DE and E detectors (as seen in Figure \ref{fig:hiralayout}) so that alpha particles are emitted directly onto the E detector without having to disassemble the array. The HiRA telescope was designed so that the pin mounted in the frame sits 3.2 mm above the surface of the E detector when the frame is inserted into the slot. Since the $^{212}$Pb on the pin source is far down the decay chain of $^{228}$Th, only two peaks feature prominently in the observed energy spectrum (as shown in the green box in Figure \ref{fig:decays}) \cite{ENSDF}. The collected data can then be used as a standalone calibration or to validate the $^{228}$Th calibrations at the end of an experiment. The latter option is in general preferable since the $^{228}$Th data have more peaks to use as calibration points.

\begin{figure}
\centering
\includegraphics[width=0.7\textwidth]{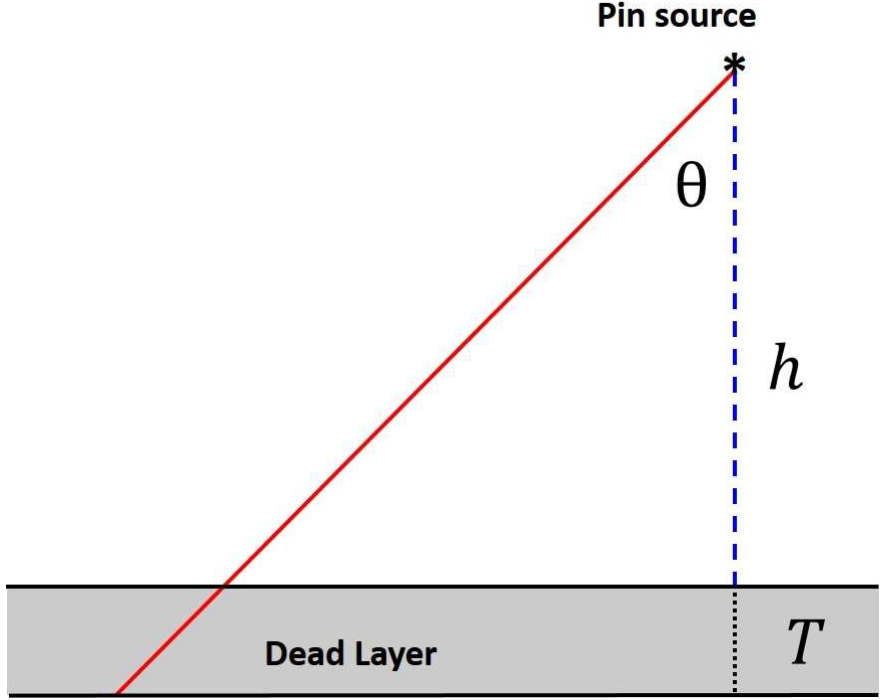}
\caption{Cartoon of pin source and dead layer. The distance $h$ between the pin source and the detector surface is 3.2 mm. The angle $\theta$ of the alpha particle from the pin source determines how much of the dead layer the particle passes through. The dead layer thickness ($T$) can be extracted by studying the relationship between detected energy and angle.}
\label{fig:pinCartoon}
\end{figure}

Of particular importance in this work is that the pin source also provides another ingredient to a proper energy calibration: determination of the thickness of the dead layer. The pin source is nestled between the DE and the E, so alpha particles from the pin do not need to pass through a gold window or Mylar foil to reach the detector (as is the case with alpha particles from the thorium source). Therefore the dead layer is the only potential cause of energy loss between the source and the active detector volume, as shown in the cartoon in Figure \ref{fig:pinCartoon}. The calibration is performed under vacuum, and the thickness of the $^{212}$Pb deposition can be neglected. Furthermore, the pin itself is only 3.2 millimeters above the detector surface, so alpha particles will travel through the dead layer at a wide range of incident angles. Although the dead layer can contain a variety of materials, we assume an effective dead layer of pure silicon since our only concern is the resulting effect on charged particles. Assuming a uniform dead layer and a constant value for the energy loss, simple geometrical considerations (illustrated in Figure \ref{fig:pinCartoon}) dictate that the energy of the alpha particles detected at different pixels can be used to extract the dead layer thickness via the following relation: 
\begin{equation}
\label{eq:dl}
E = E_{0} - \frac{dE}{dx}*\frac{T}{\cos{\theta}}
\end{equation}
where $E$ is the detected energy, $E_0$ is the initial energy of the alpha particle, $\theta$ is the emission angle, $\frac{dE}{dx}$ is the stopping power of an alpha particle in silicon at energy $E_0$ (since the dead layer is so thin, the stopping power is assumed to be constant), and $T$ is the dead layer thickness. Since the amount of dead layer that the alpha particle traverses depends on the incident angle of the particle, the final energy of that alpha particle will also depend on the incident angle. Measuring the relationship between incident angle and detected energy allows for extraction of $T$. 

\subsection{Analysis Details}
\begin{figure}
\centering
\includegraphics[width=0.7\textwidth]{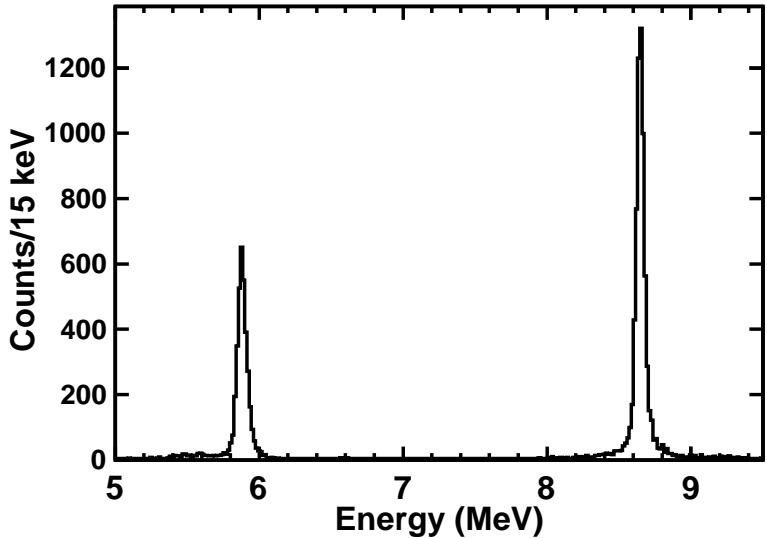}
\caption{Example pin source alpha spectrum for one pixel in one E detector.}
\label{fig:examplePb}
\end{figure}

The relevant decays for the pin source inserted into HiRA are shown in the green dashed box in Figure \ref{fig:decays}. An example spectrum in Figure \ref{fig:examplePb} illustrates that there are only two peaks in the pin source data: one that corresponds to an 8.785 MeV decay and a lower energy peak that actually consists of two unresolved alpha decay peaks (one at 6.051 MeV and another at 6.090 MeV). The relative probabilities of these two unresolved decays are well known, so these two energies can be combined via a weighted sum to yield a single peak energy of 6.062 MeV. These are not the peak energies that are seen in the detector: alpha particles will lose energy in the dead layer so the measured energies will be below the decay values. Fortunately, these deviations are exactly what allow for measurement of the dead layer thickness.

First, an initial calibration was done using the thorium source data with a reasonable guess for the dead layer thickness (on the order of 1.0 $\mu$m based on previous studies). Since the dead layer thickness is not yet precisely known, this calibration is inexact. In the fits described below, the initial energy of the alpha particle is a free parameter in order to account for this potential imperfection. 

\begin{figure}
\centering
\includegraphics[width=0.7\textwidth]{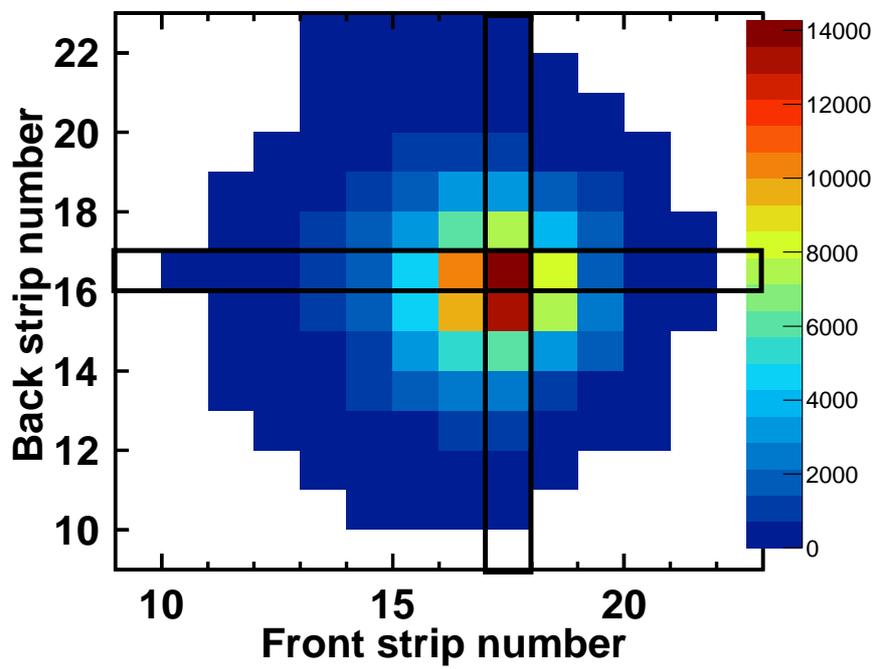}
\caption{A two-dimensional hit map for the number of detected pin source counts for one E detector. The central pixel is shown at the intersection of the black rectangles (which correspond to the front and back strips with the most counts).}
\label{fig:hitmap}
\end{figure}

The next step in determining the dead layer thickness is to find the central pixel, which is the pixel that the pin source is closest to. In principle the pin source should be exactly at the center of the detector (which falls in between pixels), but in practice the pin source could be slightly misaligned and is slightly closer to one pixel than the rest. The central pixel provides a good first approximation to the precise location of the pin source. Because of its relatively large solid angle coverage, the central pixel should have more counts than any other pixel. Figure \ref{fig:hitmap} shows a two-dimensional hit map in which the back strip axis is along the y-direction and the front strip axis is along the x-direction. The central pixel (which by definition is at the intersection of the central front strip and the central back strip) has the most counts, and the counts decrease as distance from the central pixel increases. The hit map is not perfectly concentric due to asymmetries in the deposition on the pin head, but this is a negligible effect.

\begin{figure}
\centering
\includegraphics[width=0.7\textwidth]{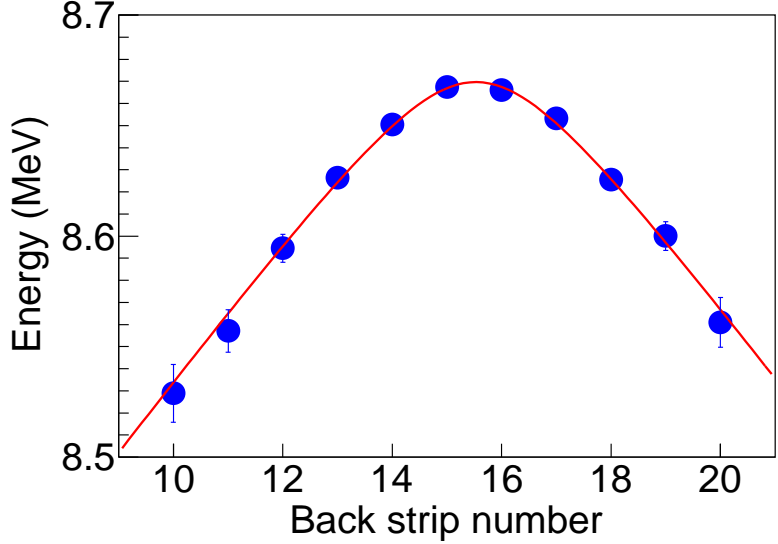}
\caption{Example energy distribution with statistical error bars of the 8.785 MeV peak for different pixels across the central front strip. The corresponding fit (using Equation 2) is shown in red.}
\label{fig:1dFit}
\end{figure}

Next, fits were performed for each peak of each pixel across the central front strip to determine precisely the measured energies for each pixel made by the intersection of the central front strip with a back strip. According to Equation \ref{eq:dl}, the central pixel should measure the highest detected energy, since the incident angle of the alpha particle is closer to 0 than for any other pixel. Figure \ref{fig:1dFit} shows the energies for several pixels across the central front strip for an example detector, as well as a fit performed using a modified version of Equation \ref{eq:dl}:
\begin{equation}
\label{eq:dl1}
E(s_{b}) = E_{0} - \frac{dE}{dx}*T*\sqrt{1 + \Big( \frac{d_0}{h} \Big) ^2 * (s_{b0} - s_{b})^2 }
\end{equation}
where $d_0$ is the width of each strip (fixed to 1.95 mm), $h$ is the distance from pin source to detector surface (fixed to 3.2 mm), and $s$ is the back strip number for each pixel. The back pin source position $s_{b0}$ refers to the exact coordinate of the pin source along the back strip axis. This position can take fractional values, e.g. $s_{b0}$ = 15.5 would indicate that the pin source is located above the space in between back strips 15 and 16. The free parameters are $T$, $s_{b0}$, and $E_0$: $E_0$ is treated as a free parameter since the dead layer thickness is not known in the initial $^{228}$Th calibration, so $E_0$ may be slightly off. A similar fit can be performed across the central back strip in order to find the front pin source position $s_{f0}$, defined similarly to $s_{b0}$.

\begin{figure}
\centering
\includegraphics[width=0.7\textwidth]{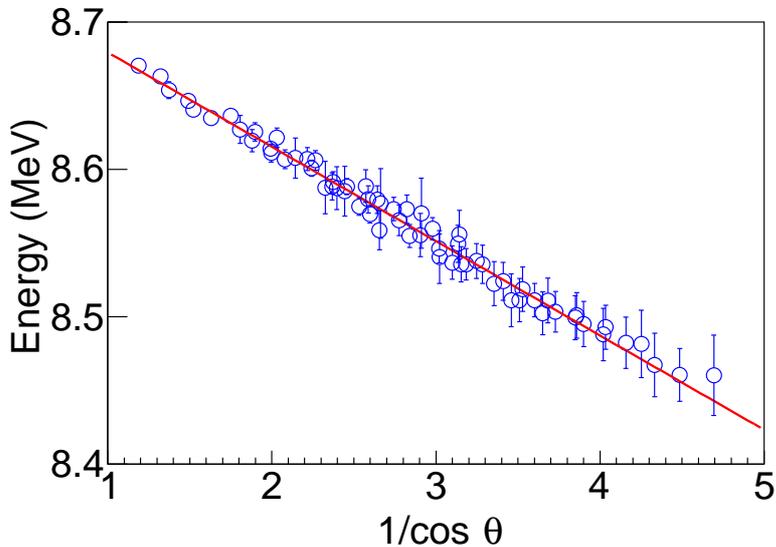}
\caption{Example energy distribution with statistical error bars for the 8.785 MeV peak across all pixels in a single telescope according to different values of $1/\cos{\theta}$, as well as the corresponding fit. }
\label{fig:2dFit}
\end{figure}

\begin{figure}
\centering
\includegraphics[width=0.7\textwidth]{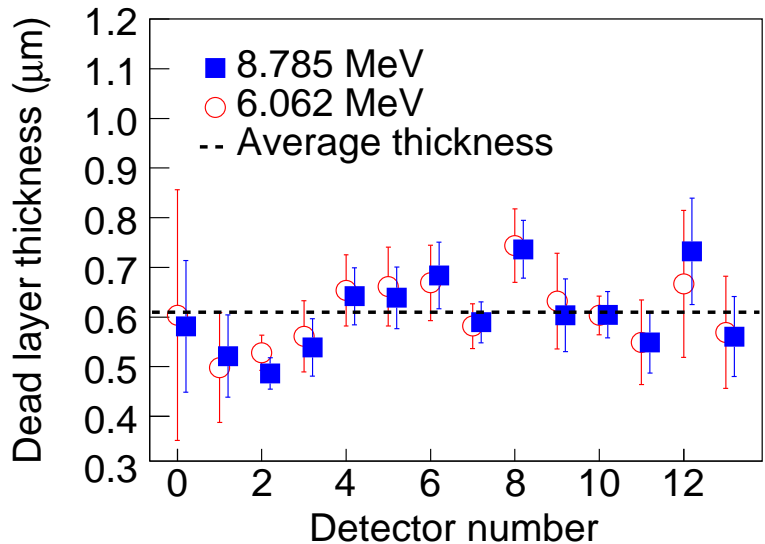}
\caption{Dead layer thicknesses extracted for each E detector from both the low (red open points) and high (blue solid points) energy peaks from $^{212}$Pb decay. All error bars are statistical, and the wide range of errors is due to different intensities of the pin sources. The average dead layer thickness (0.61 $\mu$m) is given by the dotted black line. Values extracted with the higher energy peak and the lower energy peak are consistent with each other.}
\label{fig:dlResult}
\end{figure}

Equation \ref{eq:dl1} can straightforwardly be extended to two dimensions resulting in Equation \ref{eq:dl2}, in which case the number of data points increases since the fit is no longer limited to a single central strip:
\begin{equation}
\label{eq:dl2}
E(s_{f},s_{b}) = E_{0} - \frac{dE}{dx}*T*\sqrt{1 + \Big( \frac{d_0}{h} \Big) ^2 * [ (s_{f0} - s_{f})^2 + (s_{b0} - s_{b})^2 ] }
\end{equation}
where $s_{f}$ and $s_{b}$ are the front and back strip numbers of the pixel, and $s_{f0}$ and $s_{b0}$ come from the one-dimensional fits described above. The free parameters in this fit are only $T$ and $E_0$. Results of an example fit are shown in Figure \ref{fig:2dFit}. The quality of the fit shows that our assumption of dead layer uniformity is valid, at least for the central area of the detector defined by the central 10 to 12 front strips and the central 10 to 12 back strips. The extracted dead layer thicknesses for 14 telescopes are shown in Figure \ref{fig:dlResult}, with results for both the 8.785 MeV peak (blue square symbols) as well as the 6.062 MeV peak (red open circles). The error bars are statistical uncertainties that mainly depend on the intensity of the pin source for a given telescope. Since many pins were bundled together during the source irradiation, the $^{212}$Pb isotopes could be distributed unevenly across all pins. The mean dead layer value across all telescopes of 0.61 $\pm$ 0.07 $\mu$m is indicated by the dotted line. Within error, this average matches the value provided by the manufacturer of 0.5 $\mu$m \cite{Micron}.

\section{Si Detector Thickness Determination}
\subsection{Experimental Details}

While the dead layer of the E detector influences the energy of particles that stop in the E, it has a smaller effect on particles with enough energy to punch through the entire E detector. In the latter case, the overall thickness of the E plays an important role in understanding the energetics of these punch-through particles. The CsI are typically calibrated using particles that punch-through the E. Therefore the thickness of the E detector is an important quantity for proper calibration of the CsI crystals.

\begin{figure}
\centering
\includegraphics[width=0.7\textwidth]{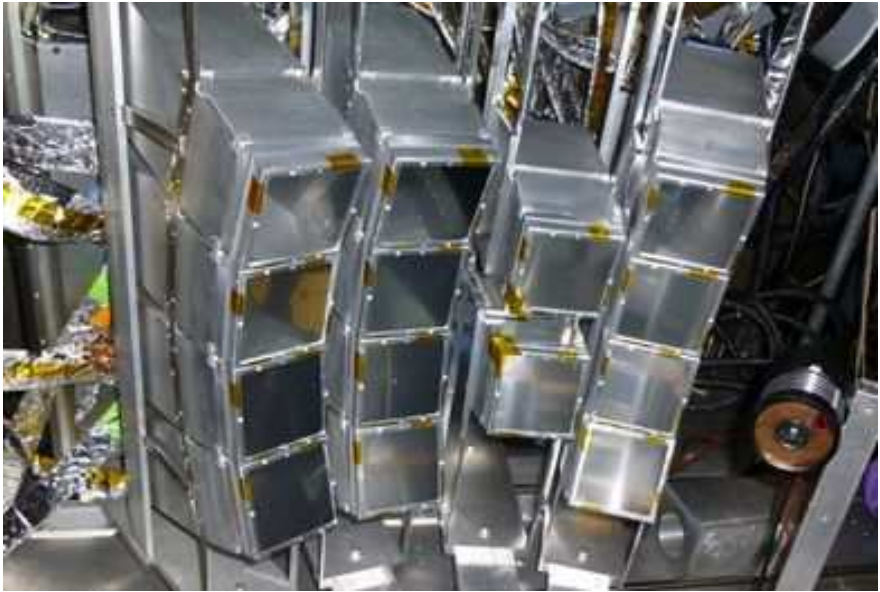}
\caption{The High Resolution Array (HiRA). In this picture, HiRA is arranged at forward angles for a transfer reaction measurement.}
\label{fig:hiraPic}
\end{figure}

\begin{figure}
\centering
\begin{minipage}{.5\textwidth}
  \centering
  \includegraphics[width=\linewidth]{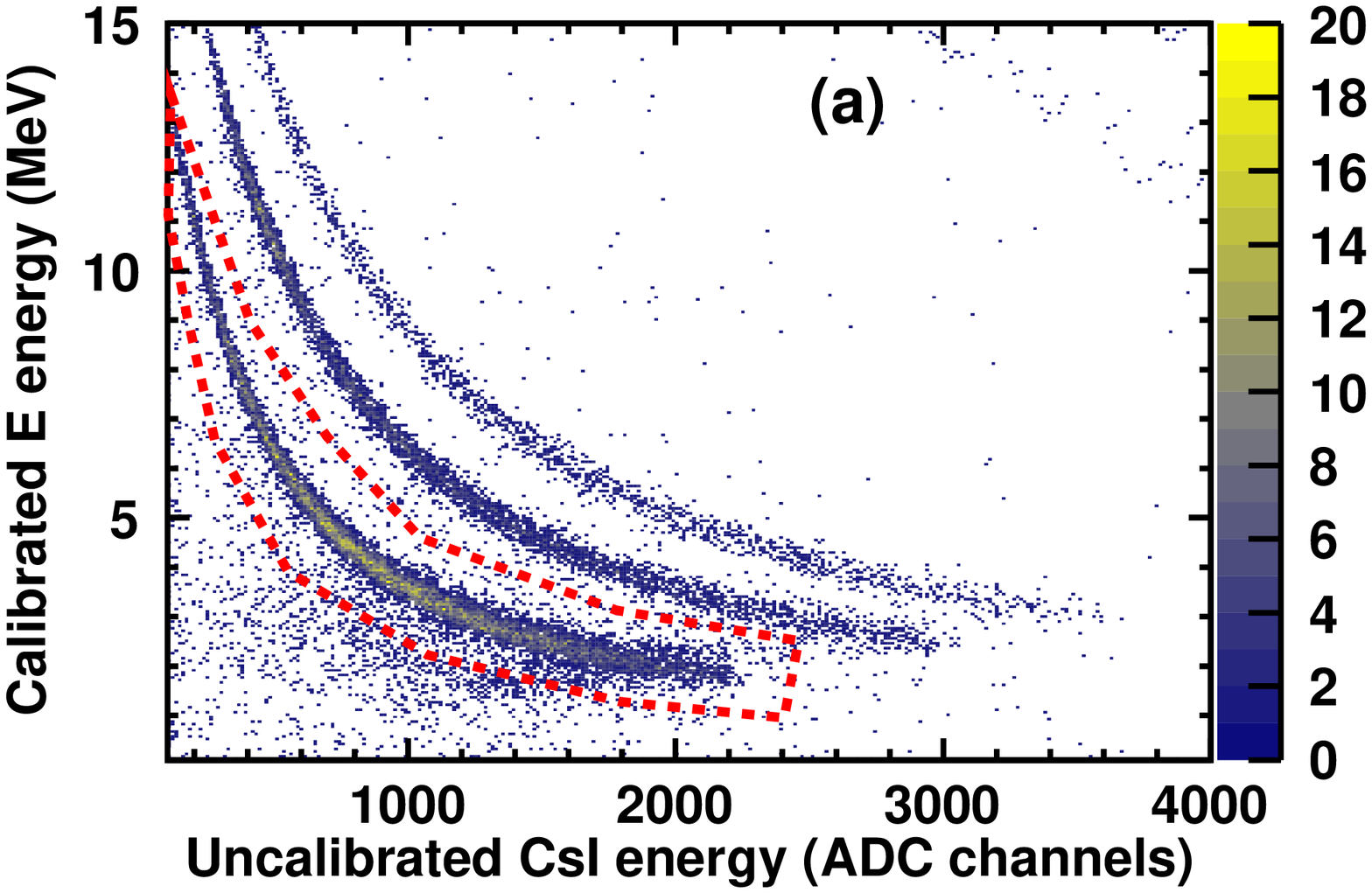}

\end{minipage}%
\begin{minipage}{.5\textwidth}
  \centering
  \includegraphics[width=\linewidth]{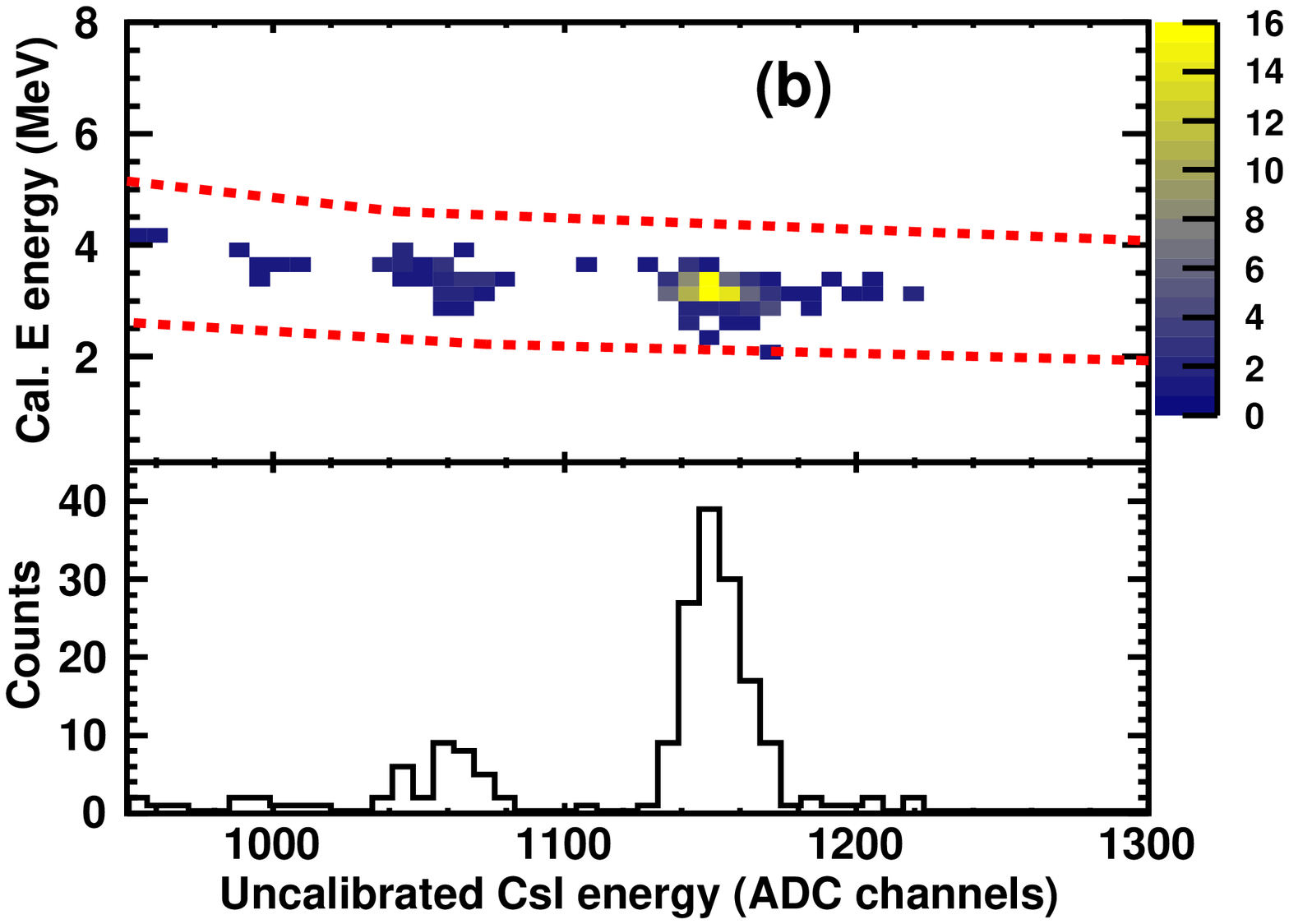}
 
\end{minipage}
 \caption{(a) HiRA PID plot for reaction data using one CsI crystal and its corresponding E silicon detector. The y-axis is calibrated energy in the E detector, and the x-axis is uncalibrated CsI energy in units of the electronics channels. Protons (within the red-dashed line) can clearly be identified, even though the CsI energy is uncalibrated. (b) The top panel is the HiRA PID for the same CsI crystal as in (a), this time showing the scattering data (zoomed in to the relevant region). The red-dashed line is the same proton gate as in (a). The bottom panel is a projection of the top panel onto the x-axis. Two peaks can clearly be seen: the higher energy peak corresponds to proton elastic scattering off of carbon, and the lower energy peak is from inelastic scattering ($E^{*}$($^{12}$C) = 4.439 MeV).}
\label{fig:hiraPID}
\end{figure}

One way to calibrate a CsI crystal is to accelerate and elastically scatter light-charged-particles at known energies into the crystal. For this work, a beam of hydrogen isotopes with a magnetic rigidity of 1.10 Tm (corresponding to a proton energy of 56 MeV) was used at the National Superconducting Cyclotron Laboratory on the campus of Michigan State University. Only protons were considered in this analysis due to low intensities for the deuteron and triton beams. HiRA was set up for a transfer reaction experiment with a configuration as shown in Figure \ref{fig:hiraPic}, covering angles from 5 to 40 degrees in the laboratory frame. A CH$_2$ reaction target (75 $\mu$m thick) scattered the incoming light charged particles. For the most part, these particles scatter off the carbon in the target. From two-body kinematics, the energies of the elastically scattered particles at a given angle are known, and range from approximately 54 to 56 MeV. These energies are high enough to punch through both the DE and the E detectors easily (the proton punch-through energies for the DE and E detectors are 2.45 MeV and 15.6 MeV, respectively), while also stopping within the CsI, which has a punch-through energy upwards of 110 MeV. Since the kinematic relationship between scattering angle and energy is relatively flat (in other words, the energy is only weakly dependent on scattering angle), this results in one elastically scattered proton calibration point per crystal, as shown in Fig. \ref{fig:hiraPID}. $^{12}$C has an excited state at 4.439 MeV, so there is an inelastically scattered proton calibration point as well. Together, these elastically and inelastically scattered protons constrain the calibration at high energy for a given crystal. There is some scattering off the hydrogen in the plastic target, but due to the low statistics and a steep kinematic relationship (compared to proton-carbon scattering) this data was not used. 

The scattering data constrain the calibration only at high energy. The calibration of the CsI at low energies utilizes the Bethe-Bloch formula by comparing energy lost by a charged particle punching through the E detector to the energy deposited in the CsI where the particle stops {\cite{Seamster77}. If we know the E detector thickness, we can calculate the energy deposited by a proton in the E detector ($E_{\mathrm{Si}}$) for a range of incoming-proton energies ($E_{p}$) using energy-loss tables {\cite{Ziegler10}}. The energy deposited in the CsI crystal is therefore $E_{\mathrm{CsI}}$ = $E_{p} - E_{\mathrm{Si}}$. $E_{\mathrm{CsI}}$ can then be used to calibrate the raw CsI ADC channels corresponding to the calibrated E energy as seen in Fig. {\ref{fig:hiraPID}}.}

This procedure (which will be referred to as the energy loss method) allows for a calibration that extends well into the low-end of the dynamic range of the CsI. However, proper implementation requires precise knowledge of the thickness of the E detector since it is a critical ingredient in determining the incoming-proton energy that corresponds to the energy deposited in the E (and therefore the calculated CsI energy used to perform the calibration). Although the nominal thickness of each E detector is 1500 $\mu$m, the true value for the thickness can differ from this value by up to 100 $\mu$m. 

\begin{figure}
\centering
\includegraphics[width=0.7\textwidth]{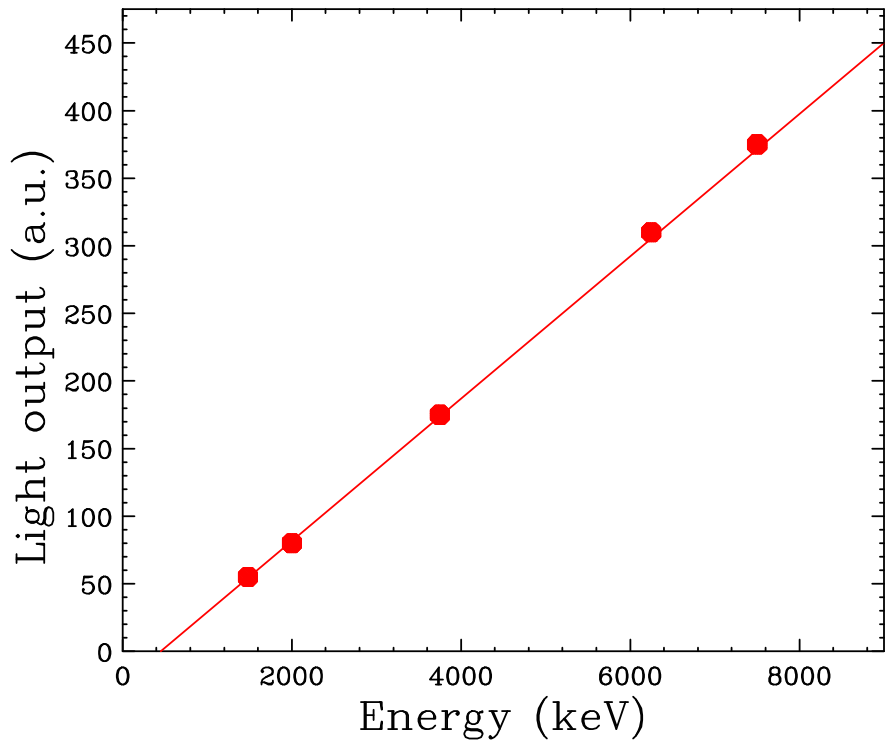}
\caption{HiRA CsI light response with a direct proton beam at low energies. Data for one crystal is shown with a linear fit. Modified from \cite{Sanetullaev12}.}
\label{fig:lightResponse}
\end{figure}

Two important notes must be made concerning the validity of this approach. First, the energy loss method relies on the assumption that the CsI detector response is linear at low energies. To confirm this, HiRA crystals were tested via direct proton beam at Western Michigan University using several different beam energies (see Figure \ref{fig:lightResponse}). The crystals were found to be linear down to approximately 1 MeV \cite{Sanetullaev12}. Secondly, since the CsI light output depends on the detected particle species, the energy loss method requires that only data from protons hitting the detector be used. Fortunately, by comparing single CsI crystals to their corresponding calibrated E detector, protons can be unambiguously identified in the CsI crystal even without a calibration (see Figure \ref{fig:hiraPID}).

\subsection{Analysis Details}

Two separate CsI calibration methods in two different energy regions were combined in order to determine the thickness of the E silicon detector within each telescope. The first calibration method is simply to use protons scattered from the CH$_2$ reaction target. The kinematics are well understood, so at a given angle the proton energy is known. These protons are at relatively high energies. As a result, they do not deposit a large amount of energy in the E, and they do not have high sensitivity to the E detector thickness; i.e. a large change in the detector thickness will only slightly change the energy lost by a high energy proton. However, the proton energy is so high that there is a long ``lever arm'' when extrapolating down to low energy. Small deviations in the high energy points will have a major impact on the low end of the dynamic range.

The second calibration method is the energy-loss method described above, which allows for a series of calibration points at low energy to be generated for a given detector thickness. These points are highly sensitive to the E thickness. The correct detector thickness should result in consistency between the low energy points calculated with the energy-loss method and the high energy scattering points. In this energy range the detector response is linear, so therefore the energy-loss and scattering points should be collinear. To check this, the energy-loss and scattering points were calculated using detector thicknesses from 1400 $\mu$m to 1600 $\mu$m. 

\begin{figure}
\centering
\includegraphics[width=0.99\textwidth]{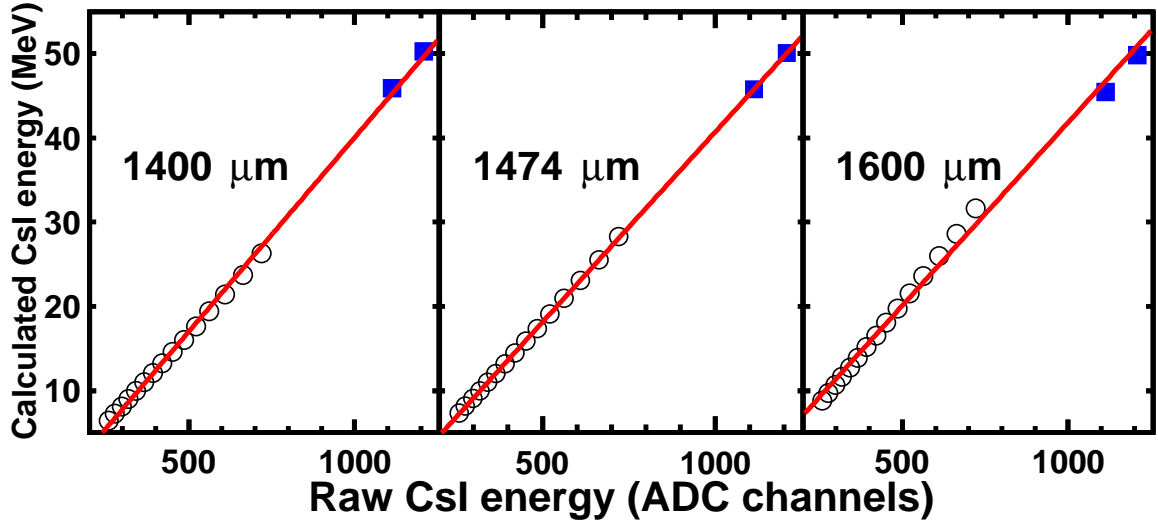}
\caption{Relationship between calculated CsI energy and raw CsI channels for one telescope at three different E thicknesses. The blue squares correspond to scattered protons with well known energies that deposit a small amount of their energy in the E detector, and the open circles are calculated via the energy-loss method as described in the text using each of the three indicated E thicknesses. When fitting both sets of points together, an E thickness of 1474 $\mu$m provides the points that yield the best fit.}
\label{fig:varyThickness}
\end{figure}

\begin{figure}
\centering
\includegraphics[width=0.7\textwidth]{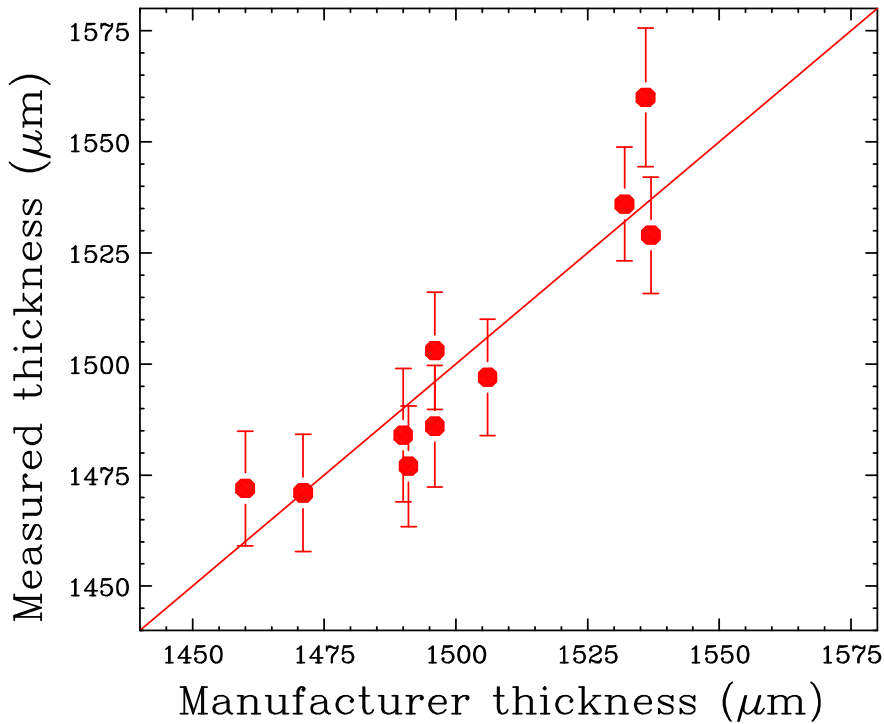}
\caption{Summary of results from E thickness extraction in comparison with manufacturer provided detector thicknesses. The line corresponds to exact agreement. The measured values agree within error to the manufacturer values. Wide variation from the nominal value of 1500 $\mu$m is evident.}
\label{fig:thicknessResults}
\end{figure}

Figure \ref{fig:varyThickness} shows example plots of CsI energy vs raw CsI channels for one crystal, with CsI energies calculated using three different values for the E detector thickness: 1400 $\mu$m, 1474 $\mu$m, and 1600 $\mu$m. The blue squares are the scattering data, and the open circles were calculated via the energy-loss method. All points were calculated using the indicated thickness. At the correct thickness value, these points should be collinear. As the assumed thickness value diverges from the correct value, the fit quality drops. Thus, the thickness can be extracted from finding the fit with minimum Chi-square. The resulting thicknesses for 10 telescopes are directly compared to the thicknesses provided by the manufacturer in Figure \ref{fig:thicknessResults}.

\section{Conclusion}

This paper presents characterization of thick (approximately 1500 $\mu$m) silicon detectors by the determination of the effective silicon dead layer thickness as well as the overall detector thickness. We show how the long-lived $^{228}$Th source can be used to calibrate silicon detectors and how the custom-made $^{212}$Pb source deposited at the tip of a pin is used to determine the effective dead layer on the silicon detector. In principle, the pin source can also be used to calibrate the E detectors placed behind a DE detector. For particles with high enough energy to punch through the E detector into the CsI detector, the E detector thickness then becomes important in determining the energetics for the calibration of the CsI detectors especially for low energy particles. We describe two methods to calibrate the CsI detectors at the high and low energy regions. By requiring collinearity of the calibrated points at the high and low energy regions, we can determine the detector thicknesses. Both the dead layer and the detector thicknesses match, within error, the values specified by the manufacturer.

\section{Acknowledgments}

This work was supported by the National Science Foundation under grant number PHY-1565546. This material is based upon work supported by the U.S. Department of Energy, Office of Science, Office of Nuclear Science, under Award Number DE-FG02-94ER40848. J.M. was supported by the Department of Energy National Nuclear Security Administration Stewardship Science Graduate Fellowship program, which is provided under grant number DENA002135.


\bibliographystyle{unsrt}

\bibliography{main}

\end{document}